\documentclass{svjour2}                    
\smartqed  
\usepackage{graphicx}
\begin{document}

\title{Implications of Space-Time foam for Entanglement Correlations of Neutral Kaons
}

\titlerunning{Implications of Space-Time foam}        

\author{Sarben Sarkar
}


\institute{Sarben Sarkar \at
              Department of Physics, King's College London, Strand, London WC2R 2LS, UK \\
              Tel.: +44-(0)20-78482514\\
              Fax: +44-(0)20-78482420\\
              \email{sarben.sarkar@kcl.ac.uk}           
}

\date{Received: date / Accepted: date}

\maketitle

\begin{abstract}
The role of $CPT$ invariance and consequences for bipartite entanglement of neutral (K) mesons are discussed. A relaxation of $CPT$ leads to a modification of the entanglement which is known as the $\omega$ effect. The relaxation of assumptions required to prove the $CPT$ theorem are examined within the context of models of space-time foam. It is shown that the  evasion of the EPR type entanglement implied by $CPT$ (which is connected with spin statistics) is rather elusive. Relaxation of locality (through non-commutative geometry) or the introduction of decoherence by themselves do not lead to a destruction of the entanglement. So far we find only one model which is based on non-critical strings and D-particle capture and recoil that leads to a stochastic contribution to the space-time metric and consequent change in the neutral meson bipartite entanglement. The lack of an omega effect is demonstrated for a class of models based on thermal like baths which are generally considered as generic models of decoherence.
\end{abstract}

\section{Introduction}
\label{intro}
It is established that the laws of Nature do not satisfy, $CP$, $PT$, and $CT$ symmetry where the operators $C$, $P$, and $T$ denote charge conjugation, parity and time reversal respectively \cite{Lee}. Currently the successful theories are based on local Lorentz invariant lagrangians, and it has been shown given the spin-statistics connection that $CPT$ is a symmetry. The symmetry implies that the solution set of a theory is invariant under reversal of parity, time and  interchange of particle and antiparticle. Consequently any violations of the consequences of $CPT$ symmetry \cite{streater} would entail physics beyond the standard model of particle physics which is based on lagrangians. Typically the consequences of $CPT$ that are considered are those of equal masses and lifetimes for particles and antiparticles. Recently it was noted \cite{Bernabeu} that when the $CPT$ operator is not well
defined there are implications for the symmetry structure of the initial
entangled state of two neutral mesons in meson factories such as DA$\Phi$NE, the Frascati $\phi $ factory. Indeed, if $CPT$ can
be defined as a quantum mechanical operator, then the decay of a (generic)
meson with quantum numbers $J^{PC}=1^{--}$ \cite{Lipkin}, leads to a pair
state of neutral mesons $\left\vert i\right\rangle $ having the form of the
entangled state%
\begin{equation}
\left\vert i\right\rangle =\frac{1}{\sqrt{2}}\left(  \left\vert \overline
{M_{0}}\left(  \overrightarrow{k}\right)  \right\rangle \left\vert
M_{0}\left(  -\overrightarrow{k}\right)  \right\rangle -\left\vert
M_{0}\left(  \overrightarrow{k}\right)  \right\rangle \left\vert
\overline{M_{0}}\left(  -\overrightarrow{k}\right)  \right\rangle \right)  .\label{CPTV}
\end{equation}
This state has the Bose symmetry associated with particle-antiparticle
indistinguishability $C\mathcal{P}=+$, where $C$ is the charge conjugation and
$\mathcal{P}$ is the permutation operation. If, however, $CPT$ is not a good
symmetry (i.e. ill-defined due to space-time foam), then $M_{0}$ and
$\overline{M_{0}}$ may not be identified and the requirement of $C\mathcal{P}%
=+$ is relaxed~\cite{Bernabeu}. Consequently, in a perturbative framework, the
state of the meson pair can be parametrised to have the following form:
\[
\left\vert i\right\rangle    =\frac{1}{\sqrt{2}}\left(  \left\vert
\overline{M_{0}}\left(  \overrightarrow{k}\right)  \right\rangle \left\vert
M_{0}\left(  -\overrightarrow{k}\right)  \right\rangle -\left\vert
M_{0}\left(  \overrightarrow{k}\right)  \right\rangle \left\vert
\overline{M_{0}}\left(  -\overrightarrow{k}\right)  \right\rangle \right)
\\
  +{\frac{\omega}{\sqrt{2}}}\left| {\Delta \left( {\vec k} \right)} \right\rangle \label{omegaterm}
\]
where
\[\left| {\Delta \left( {\vec k} \right)} \right\rangle  \equiv \left| {{{\overline M }_0}\left( {\vec k} \right)} \right\rangle \left| {{M_0}\left( { - \vec k} \right)} \right\rangle  + \left| {{M_0}\left( {\vec k} \right)} \right\rangle \left| {{{\overline M }_0}\left( { - \vec k} \right)} \right\rangle \]
and $\omega=\left\vert \omega\right\vert e^{i\Omega}$ is a complex $CPT$
violating (CPTV) parameter. For definiteness in what follows we shall term
this quantum-gravity effect in the initial state as the \textquotedblleft%
$\omega$-effect\textquotedblright\cite{Bernabeu}. There is actually another dynamical
\textquotedblleft$\omega$-effect\textquotedblright\ which is generated during
the time evolution of the meson pair and this will be discussed further in passing \cite{bernabeu}.

Gravitation is one interaction which is not included in the standard model and we shall assume that it has a quantum manifestation. We will examine whether any such manifestation implies an $\omega$-effect. Although there is no satisfactory theory of quantum gravity there are features which will most likely survive a full resolution. One current attempt is based on string/brane theory \cite{zwiebach} and will be for us a primary motivation. On more general grounds the existence of black hole solutions in general relativity and semi-classical arguments which imply thermal Hawking radiation emitted by a black hole have thrown into doubt that unitarity is preserved in quantum evolution \cite{HawkingPage}. Since the semi-classical calculation embodies the classical causality structure of horizons it is perhaps not surprising that unitarity is found not to hold. Clearly the relevance of an absence of unitarity is in the lack of inverse for T which will then have implications for $CPT$. There has been a continuing debate on the absence or otherwise of unitarity. The more recent contributions have suggested that the correlations between  the Hilbert space of states within the horizon and the outgoing radiation would allow a  recovery of the information in the black hole \cite{thooft}. However considerations from quantum information theory based on estimates of mutual information have indicated that the evolution in the presence of even evaporating black holes is non-unitary \cite{braunstein,oppenheim}. In the presence of space-time foam which can be due to formation and destruction of microscopic black holes or other space-time singularities such as D-particles from the bulk in string/brane theory impinging on a $3$ dimensional brane \cite{emw} , we feel non-unitary behaviour is likely. Furthermore in non-critical string theory there are plausible arguments based on the renomalization group which could be interpreted in terms of non-unitary time-evolution. In section 2 we shall first discuss how the existence of the $\Theta\equiv{CPT}$ is incompatible with non-unitary evolution. We will elaborate on this with some arguments favouring the existence of non-unitary evolution both on general grounds and also within the context of non-critical strings. In section 3 we will give a model inspired by string induced decoherence \cite{Sarkar} and show that the $\omega$-effect is generated. In section 4 we will evaluate plausible alternative models of space-time foam such as noncommutative geometry \cite{balachandran} and thermal baths and conclude that the omega effect is an extremely fragile phenomenon.

\section{Master Equation and $CPT$}
\label{sec:2}
The implications of non-unitary evolution for $\Theta$ were first addressed by Wald \cite{wald}. Unitary time evolution is an implicit assumption in the proof of the $CPT$ theorem and so his work serves as a conceptual cornerstone for the approach to the omega effect that we develop in this paper. A brief review of the discussion of Wald will now be given. Within the context of scattering theory let $\mathcal{H}_{in}$ denote the Hilbert space of in states and $\bar{\mathcal{H}}_{in}$ the dual space. We can define the analogous entities for the out states. Let $S$  be a mapping from the set of in-states $\mathcal{G}_{in}$ to the set of out-states $\mathcal{G}_{out}$. In our framework $\mathcal{G}_{in}$ is isomorphic to ${\mathcal{H}_{in}}\bigotimes{\bar{\mathcal{H}}_{in}}$ since the states are represented as density matrices ${{{\rho}_{in}}^{A}}_{B}$ where $A$ is a vector index associated with $\mathcal{H}_{in}$ and $B$ is a vector index associated with $\bar{\mathcal{H}}_{in}$. The indexed form of $S$ is ${{{S^{a}}_b}_C}^{D}$ where the lower case indices refer to $\mathcal{H}_{out}$ and $\bar{\mathcal{H}}_{out}$. If probability is conserved
\begin{equation}\label{prob}
    tr\left({S}\rho\right)=\left(\rho\right)
\end{equation}
which in index notation can be written as
\begin{equation}
    {{{{S^{a}}_a}_C}^{D}}={\delta_{c}}^{D}
\end{equation}
where we have adopted the summation convention for repeated indices. Consider now operators ${\Theta}_{in}$ and ${\Theta}_{out}$ which implement the $CPT$ transformation on $\mathcal{G}_{in}$ and $\mathcal{G}_{out}$ respectively i.e.
\begin{eqnarray}
  {\Theta}_{in}& \colon & {\mathcal{G}_{in}}\rightarrow{\mathcal{G}_{out}} \\
 {\Theta}_{out}& \colon  & {\mathcal{G}_{out}}\rightarrow{\mathcal{G}_{in}}.
\end{eqnarray}
In particular under a $CPT$ transformation let ${\rho_{in}}\;{\epsilon}\;{\mathcal{G}_{in}}$  be mapped into ${\rho_{out}}'$ and  ${\rho_{out}}\;{\epsilon}\;{\mathcal{G}_{out}}$ be mapped into ${\rho_{out}}'$.
If the theory (including quantum gravity) is assumed to be invariant under $CPT$ then
\begin{eqnarray}
{\rho_{out}} &=& S{\rho_{in}},\label{CPT1} \\
{\rho_{out}}'&=& S{\rho_{in}}'\label{CPT2}\\
{{\Theta}_{in}}{\rho_{in}}&=&{\rho_{out}}',\label{scat1}\\
{{\Theta}_{out}}{\rho_{out}}&=&{\rho_{in}}',\label{scat2}
\end{eqnarray}
and
\begin{equation}\label{compos}
    {{\Theta}_{in}}{{\Theta}_{out}}=I, \; {{\Theta}_{out}}{{\Theta}_{in}}=I.
\end{equation}
Since
\begin{equation}
    {{\Theta}_{out}}={{\Theta}_{in}}^{-1}
\end{equation}
 it is convenient to drop the suffix $in$ in this last relation. Hence from (\ref{CPT1}),(\ref{CPT2}),(\ref{scat1}) and (\ref{scat2}) we can deduce that
\begin{eqnarray}
  {\Theta}{\rho_{in}} &=& {\rho_{out}}' ,\label{scat3} \\
  &=& S{\rho_{in}}' \\
  &=& S{\Theta^{-1}}{\rho_{out}}\\
  &=& S{\Theta^{-1}}S{\rho_{in}},\label{inv1}.
\end{eqnarray}
From (\ref{scat3}) and (\ref{inv1}) we can deduce the important result that $S$ has an inverse given by ${\Theta^{-1}}S{\Theta^{-1}}S$. Hence if $\Theta$ exists then time reversed evolution is permitted. Consequently for non-unitary evolution $\Theta$ cannot be defined.

\subsection{Non-unitary evolution}
\label{sec:2.1}
Given the lack of a full theory of quantum gravity it may seem that we cannot say anything rigorous about the issue of non-unitarity when quantum effects and gravity are significant. This lack of a theory has permitted various suggestions which have typically had a certain speculative element.  We will nonetheless examine some clues from the current understanding. One finding based on semi-classical analysis is thermal radiation produced by evaporating black holes which are predicted by general relativity. Thermal radiation carries no information about the initial state of the black hole and hence there is a loss of information. Loss of information is related to non-unitary evolution in the presence of black holes \cite{HawkingPage}. This was the intriguing possibility suggested (quite sometime ago) by this analysis. Clearly these (semi-classical) considerations become more and more suspect when the black hole approaches the size of the Planck scale (based on heuristic arguments combining quantum mechanics and general relativity). More recently it was suggested that information is actually not lost but remains in a stable remnant of the size of the Planck scale. This is problematic because the entropy of a black hole (admittedly based on semi-classical analysis) is proportional to the square of its mass \cite{bekenstein,hawking}. Hence a Planck sized black hole has a much smaller mass $M_{f}$ compared to the original black hole of mass $M$ and so cannot contain the original information. If there were many remnants (of the order of $\frac{M^{2}}{{M_{f}}^2}$ accommodating the $M^2$ bits of information) then these would also have had observable consequences for low energy physics \cite{oppenheim}. A more recent suggestion considered the possibility that the information was actually locked in correlations between the emitted radiation and evaporating black hole.This is akin to quantum cryptography where the key is the small black hole. Although classically the bits  required for the key would be similar to that of the original black hole, quantum mechanics allows keys that require far less bits. Such a mechanism would give the possibility of recovering the information of the black hole from correlations with the black hoke remnant. However this possibility can be ruled out \cite{braunstein} if the Hawking radiation picture persists beyond the semi-classical analysis as follows. We first note that, for a composite system made of two parts $A$ and $B$, any pure state $|\Psi\rangle$ has the Schmidt decomposition
\begin{equation}\label{Schmidt}
    |\Psi\rangle=\sum_{s}{{\eta_{s}}{|s_{A}\rangle}{|s_{B}\rangle}}
\end{equation}
where ${\eta_{s}}\geq{0}$, $\sum_{s}{\eta_{s}}^{2}=1$ and ${|s_{A}\rangle}$ and ${|s_{B}\rangle}$ are orthonormal states for $A$ and $B$ respectively. The blackhole evaporation  process can be regarded as a mapping $\mathcal{S}$ from an initial state $\varrho_{j}$ in a space $\mathcal{I}$ to a state $\rho_{f}$ in a space $\mathcal{O}$; $\rho_{f}$ is independent of the states in $\mathcal{I}$. This structure formalises the picture obtained for black hole evaporation and Hawking radiation from semi-classical analysis.The remainder of the final space can be denoted by $\mathcal{F}$, and within a quantum information context is labelled the ancilla space. If $\varrho_{j}$  is taken to be ${\left| \Psi  \right\rangle _I}{}_I\left\langle \Psi  \right|$ then we can write $\mathcal{S}{\left| \Psi  \right\rangle _I}$ , using the Schmidt decomposition, as
\begin{equation}\label{evoln}
    \mathcal{S}{\left| \Psi  \right\rangle _I}
    ={\sum_{k=1}^{K}}\sqrt{p_{k}}|k\rangle_{I}\bigotimes|F_{k}(\Psi)\rangle_{\mathcal{F}}
\end{equation}
where $\{|F_{k}(\Psi)\rangle\}$ are orthonormal states in $\mathcal{F}$, $|k\rangle$ is an eigenvector (in $\mathcal{O}$) of the map $F$ with a non-zero eigenvalue $p_{k}$; there are $K$ such eigenvalues. If $\mathcal{I}$ has dimensionality $d$ with a basis set $|\Psi_{j}\rangle$ then $\mathcal{F}$ is spanned by $\{|F_{kj}\rangle\}$ where $ \{|F_{kj}\rangle\}\equiv{|F_{k}(\Psi_{j})\rangle}\}$. Using a unitary transformation it is possible to map $\{|F_{kj}\rangle\}$ to $|q_{k}\rangle\bigotimes|{\Psi_{j}}\rangle$ where $\{|q_{k}\rangle\}$ is an orthonormal set in $\mathcal{O}$. Hence
\begin{equation}\label{evoln2}
    \mathcal{S}{\left| \Psi  \right\rangle _I}
    ={\sum_{k=1}^{K}}\sqrt{p_{k}}|k\rangle_{I}\bigotimes(|q_{k}\rangle\bigotimes|{\Psi}\rangle)_{\mathcal{F}}.
\end{equation}
Hence information concerning $|\Psi\rangle$ lies purely in $\mathcal{F}$ and not in correlations between $\mathcal{F}$ and $\mathcal{O}$ \cite{braunstein}. COnsequently the suggestion that correlations can be the key to the information problem does not seem to be viable, at least within this simple framework. It should be stated that other treatments based on extremality of black holes, anti-de Sitter
(supersymmetric) space-times and the Euclidean formulation for summing over
space-time geometries purport to support that information can be hidden (through entanglement) within correlations in a holographic way \cite{thooft}.
The recent arguments of
Hawking and other (string) theorists are somewhat special in their details  and
require constructions such as the extremality of black holes, anti-de Sitter
(supersymmetric) space-times and the Euclidean formulation for summing over
space-time geometries \cite{giddings}. Although the special properties present in the solutions can be regarded as a drawback, nevertheless they indicate that the semi-classical argument on which the earlier argument was based on may be a weak point of the analysis.We are unable to comment further on these issues and will follow arguments that we find appealing.

\subsection{Master equation from strings}
\label{sec:2.2}
Rather than just postulating non-unitary evolution based on semi-classical arguments already alluded to in general relativity we will give arguments in string theory that suggest non-unitary evolution in certain circumstances. Strings are fields over a two dimensional world sheet. The renormalization group for field theories in two  dimensions has properties which will have an interesting reinterpretation for string theories \cite{shore,zam}. In theory space, i.e. in the infinite dimensional space of couplings $\{g_{i}\}$ for all renormalizable two-dimensional unitary  quantum field theories, there exists a function $c(\{g_{i}\})$ which has the following properties:
\begin{itemize}
 \item $c(\{g_{i}\})$ is non-negative and non-increasing on renormalization group flows towards an infrared fixed point,
  \item the renormalization group fixed points are also critical points of $c(\{g_{i}\})$,
  \item value of critical value of $c(\{g_{i}\})$ is the conformal anomaly.
  \end{itemize}
More precisely in terms of a renormalization group flow parameter $t$
\begin{equation}\label{Zam}
    \frac{d}{dt}c(\{g^{i}\})=\beta^{i}G_{ij}\beta^{j}
\end{equation}
where the renormalization group $\beta$  function is defined by $\beta^{i}=\frac{d}{dt}g^{i}(t)$ and $G_{ij}$ is referred to as the Zamolodchikov metric. $G_{ij}$ is negative definite and is the matrix of second derivatives of the free energy. In the world sheet conformal field theory approach to perturbative  string theory it has long been thought that the time evolution of the string backgrounds and world sheet renormalization group flows are connected \cite{gutperle,EllisErice}. In string theory for the backgrounds to be consistent with a classical space-time interpretation the central charge will need to  have the critical values. In order to move away from such a classical background and allow in principle for fluctuating backgrounds it is necessary to deform the conformal points through vertex operators $V_{g^{i}}$ associated with background fields $g^{i}$. Hence the conformal invariant action $\emph{S}^{*}$ becomes deformed to $\emph{S}$ where
\[
    \emph{S}=\emph{S}^{*}+g^{i}\int{d^{2}z V_{g_{i}}(\Xi)}
\]
$z$ is the usual holomorphic world sheet co-ordinate, $\Xi$ are target space matter fields and $ V_{g^{i}}$ has conformal weight $(1,1)$; however (for convenience now dropping the index $i$)  $ V_{g}$ has a scaling dimension $\alpha_{g}$ to $O(g)$ with $\alpha_{g}=-g\textsl{C}_{ggg}+ \ldots$ and $\textsl{C}_{ggg}$ is the expansion coefficient in the operator product expansion of $V_{g}$ with itself. In non-critical strings conformal invariance is restored by gravitational dressing with a Liouville field $\varphi$ (which can be viewed as a local world sheet scale) so that
\begin{equation}\label{dressing}
    \int{d^{2}z g V_{g}(\Xi)}\rightarrow\int{d^{2}z ge^{\alpha_{g}\varphi}V_{g}(\Xi)}.
\end{equation}
In a small $g$ expansion
\[
\int{d^{2}z g V_{g}(\Xi)}\rightarrow \int{d^{2}z g V_{g}(\Xi)}-\int{d^{2}z g^{2}\textsl{C}_{ggg}}.
\]
Scale invariance is restored by defining a renormalized coupling $g_{R}$
\begin{equation}\label{dressing2}
 g_{R}=g-\textsl{C}_{ggg}\varphi{g^{2}}
\end{equation}
The local scale interpretation of $\varphi$ is clearly consistent with the renormalisation group $\beta$  function that we have noted earlier. The integration over world sheet metrics $\gamma_{\alpha\beta}$ ( in the Polyakov string action) implies an integration over $\varphi$ on noting that
$\gamma_{\alpha \beta}=
e^{\varphi}{\widehat{\gamma}}_{\alpha \beta}$
where ${\widehat{\gamma}}_{\alpha \beta}$ is a fiducial metric. In this way $\varphi$ becomes a dynamical variable with a kinetic term. For matter fields with central charge $c_{m}>25$ the signature of this term is opposite to the kinetic terms for the fields $\Xi$ and it has been suggested that in this case the zero mode of $\varphi$ is a target time $t$. The requirement of renormalizability of the world sheet $\sigma$ model implies that for the density matrix $\rho$ of a string state propagating in a background $\{g_{i}$
\begin{equation}\label{density}
    \frac{d}{dt}\rho(g^{i},p_{i},t)=0
\end{equation}
where $p_{i}$ is the conjugate momentum to $g^{i}$ within the framework of a dynamical system with hamiltonian $H$ and action the Zamolodchikov c-function \cite{mavspinstat}, i.e.
\begin{equation}\label{dynamical}
    c[g]=\int dt(p_{i}\dot{g}^{i}-H).
\end{equation}
From (\ref{density}) we deduce that
\begin{equation}\label{density2}
    \frac{\partial{\rho}}{\partial{t}}+\dot{g}^{i}\frac{\partial{\rho}}{\partial{g^{i}}}
    +\dot{p}_{i}\frac{\partial{\rho}}{\partial{p^{i}}}=0.
\end{equation}
The piece $\dot{p}_{i}\frac{\partial{\rho}}{\partial{p^{i}}}$ in (\ref{density2}) can be written as $G_{ij}\beta^{j}\frac{\partial{\rho}}{\partial{p^{i}}}$; using the canonical relationship of $g^{i}$ and $p^{i}$ this can be recast as $-iG_{ij}\beta^{j}[\rho,g^{i}]$. As discussed in \cite{mavspinstat}, such a term leads to a non-unitary evolution of $\rho$. However this analysis cannot be regarded as settling the issue of non-unitarity even within the framework of non-critical string theory since there are issues relating to the time like signature of the Liouville field and the identification of the local renormalization group scale. Furthermore the role of D branes and in particular D particles have not been incorporated into the analysis.

An additional reason~\cite{mavcptdecoh} for considering quantum
decoherence models of quantum gravity, comes from recent astrophysical
evidence on a current-era acceleration of our Universe. Indeed, observations
of distant supernovae~\cite{snIa}, as well as WMAP data~\cite{wmap} on the
thermal fluctuations of the cosmic microwave background (CMB), indicate
that our Universe is at present in an accelerating phase, and that 73\% of
its energy-density budget consists of an unknown substance, termed \emph{%
Dark Energy}. Best-fit models of such data include
Einstein-Friedman-Robertson-Walker Universes with a non-zero cosmological
\emph{constant}. However, the data are also currently compatible with
(cosmic) time-dependent vacuum-energy-density components, relaxing
asymptotically to zero~\cite{steinhardt}. In colliding brane-world models,
by treating the dark energy component of the Universe as a non-equilibrium
energy density of the (observable) brane world~\cite{gravanis,emw} this
density is identified with the central charge surplus of the supercritical $%
\sigma $-models describing the (recoil) string excitations of the brane
after the collision.The relaxation of the dark-energy density component can
be a purely stringy feature of the logarithmic conformal field theory~\cite%
{lcftmav} describing the D-brane recoil~\cite{kmw} in a (perturbative) $%
\sigma $-model framework.
\section{D-particle space-time foam}
\label{sec:1}
\begin{figure}[t]
\centering\includegraphics[width=7.5cm]{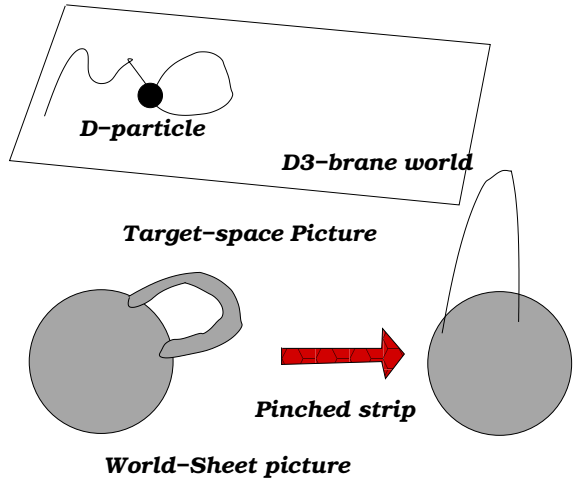}\caption{\emph{Upper
picture:} A Fluctuating D-particle is described by open strings attached to
it. As a result of conservation of fluxes~\cite{polch,polch2,johnson} that
accompany the D-branes, an isolated D-particle cannot occur, but it has to be
connected to a D-brane world through flux strings. \emph{Lower picture}:
World-sheet diagrams with annulus topologies, describing the fluctuations of
D-particles as a result of the open string states ending on them. Conformal
invariance implies that pinched surfaces, with infinitely long thin strips,
have to be taken into account. In bosonic string theory, such surfaces can be
resummed~\cite{szabo}. }%
\label{fig:dbranes}%
\end{figure}

\begin{figure}[th]
\centering\includegraphics[width=7.5cm]{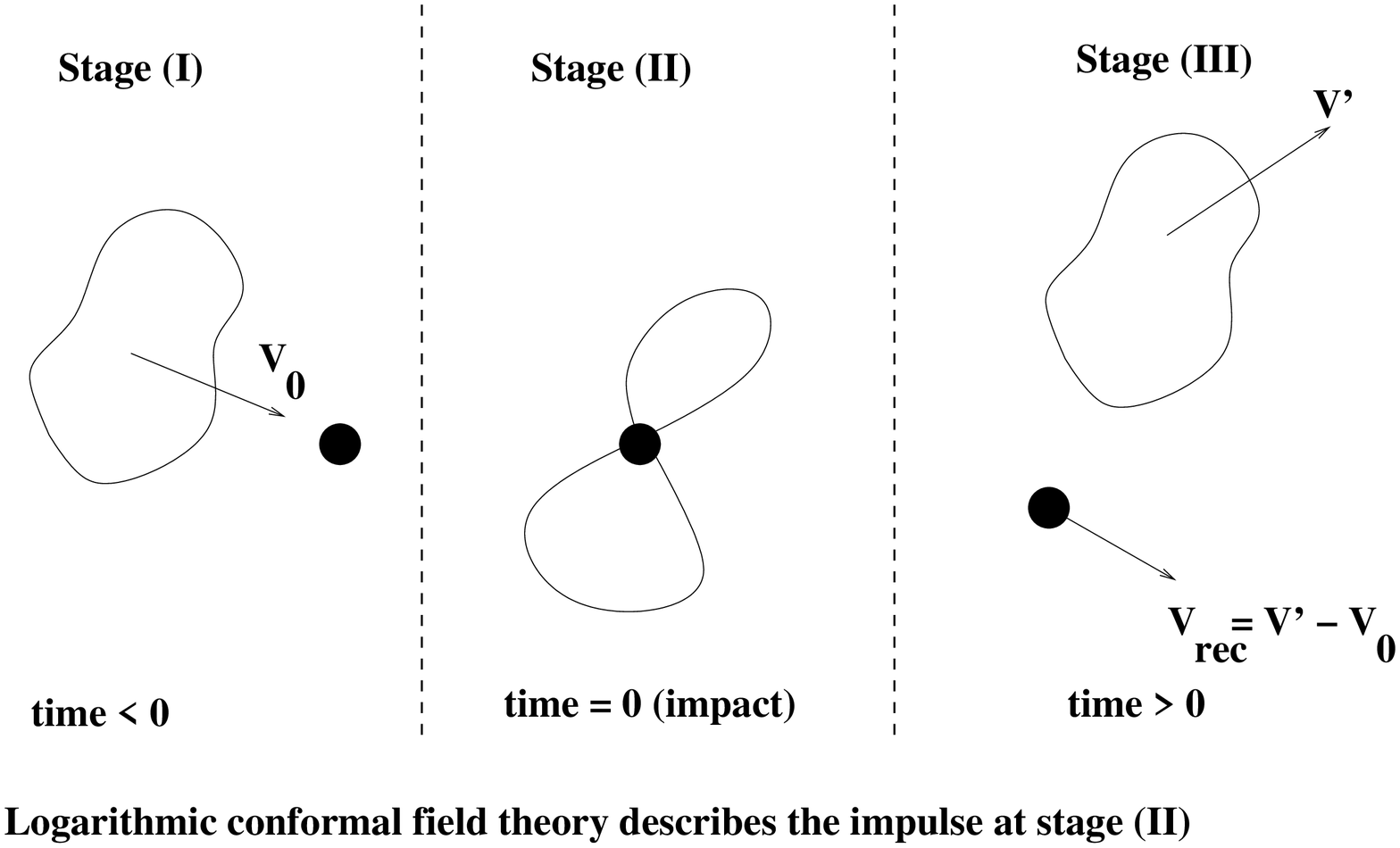} \hfill
\includegraphics[width=7.5cm]{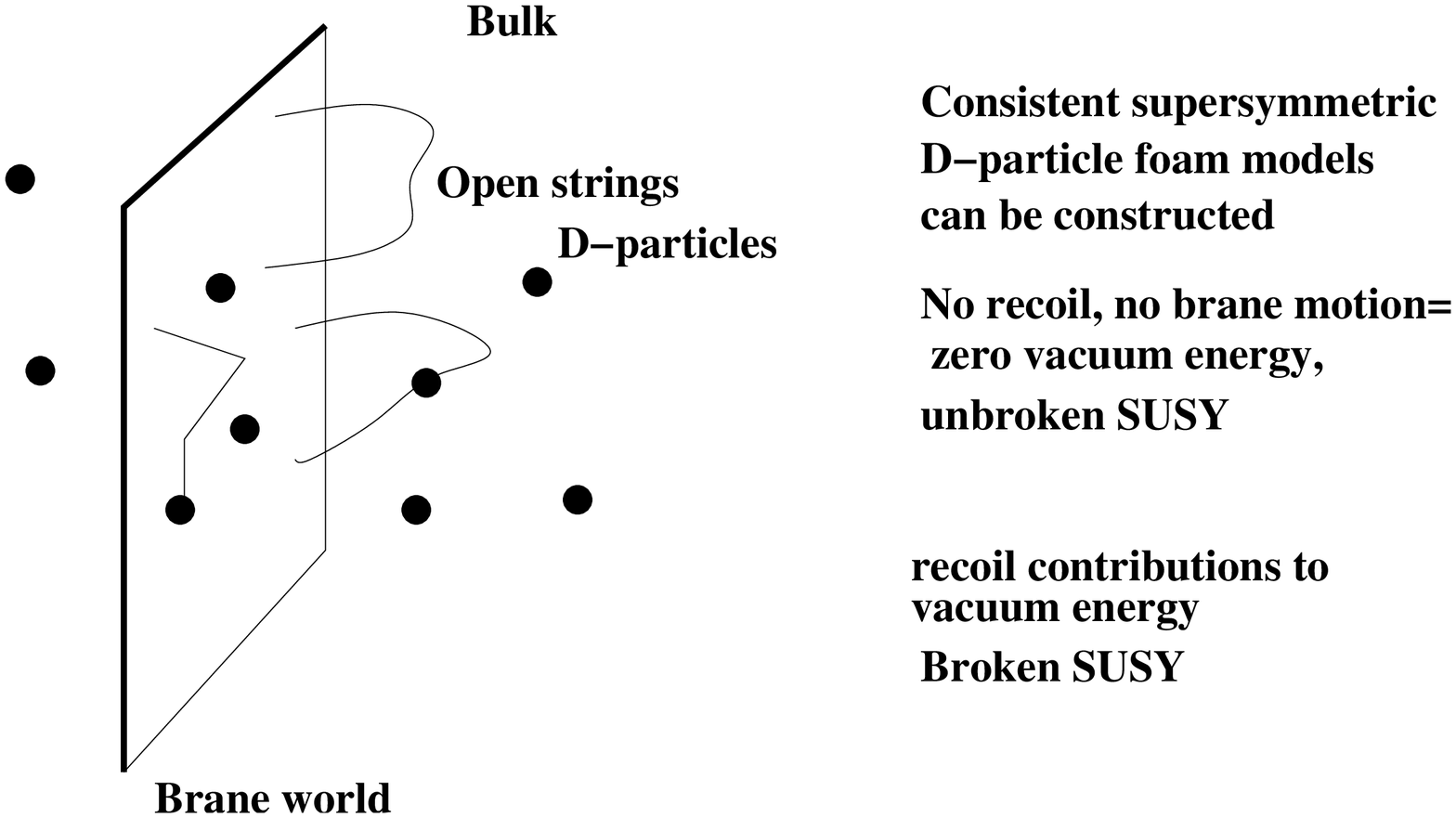} \caption{Schematic
representation of a D-foam. The figure indicates also the capture/recoil
process of a string state by a D-particle defect for closed (left) and open
(right) string states, in the presence of D-brane world. The presence of a
D-brane is essential due to gauge flux conservation, since an isolated
D-particle cannot exist. The intermediate composite state at $t=0$, which has
a life time within the stringy uncertainty time interval $\delta t$, of the
order of the string length, and is described by world-sheet logarithmic
conformal field theory, is responsible for the distortion of the surrounding
space time during the scattering, and subsequently leads to induced metrics
depending on both coordinates and momenta of the string state. This results on
modified dispersion relations for the open string propagation in such a
situation~\cite{Dfoam}, leading to non-trivial \textquotedblleft
optics\textquotedblright\ for this space time.}%
\label{fig:recoil}%
\end{figure}

D-particles are D-branes with zero spatial dimension. Instead of considering microscopic black holes as agents for the foaminess of space-time we shall consider that D-particles can play that role. Typically open strings interact with D-particles and satisfy Dirichlet boundary conditions when attached to them. Closed strings may be cut by D-particles. D-particles are allowed in certain string theories such as bosonic, type IIA and type I and here we will here consider them to be present in string theories of phenomenological interest. Furthermore even when elementary D particles cannot exist consistently there
can be effective D-particles formed by the compactification of higher
dimensional D branes. Moreover D particles are
non-perturbative constructions since their masses are inversely proportional
to the the string coupling ${g_{s}}$ . The study
of D-brane dynamics has been made possible by Polchinski's realisation that
such solitonic string backgrounds can be described in a conformally invariant
way in terms of world sheets with boundaries \cite{polch2}. On these
boundaries Dirichlet boundary conditions for the collective target-space
coordinates of the soliton are imposed \cite{coll}. Although recently a particular supersymmetric version of D-particle foam has been suggested involving stacks of $D8$ branes and $O8$ orientifold planes enclosing a bulk space \cite{emw}, it will suffice for us to consider for simplicity a model based on D-particles populating a bulk geometry between parallel
D-brane worlds. The model is termed D-foam~\cite{Dfoam} (c.f. figure
\ref{fig:recoil}), since our world is modelled as a three-brane moving in the
bulk geometry. Ordinary matter lives on the brane. As the brane moves through the bulk, stringy matter interacts with the D-particles.The spectrum of open strings attached to a Dp brane (with $p>0$ )
contains a Maxwell field and the ends of the open string carry charge. The
associated Maxwell field is confined to the world volume of the D brane. Hence
a conventionally charged string cannot end on a D-particle. We will thus
restrict our consideration to neutral particles that are \textquotedblright
captured\textquotedblright\ \ by D-particles. An important symmetry of this first quantised string
theory is conformal invariance and the requirement of the latter does
determine the space-time dimension and/or structure. \ This symmetry leads to
a scaling of the metric and permits the representation of interactions through
the construction of measures on inequivalent Riemann surfaces \cite{strings}.
In and out states of stringy matter are represented by punctures at the
boundaries. The D-particles as solitonic states~\cite{polch2} in string theory
do fluctuate themselves, and this is described by stringy excitations,
corresponding to open strings with their ends attached to the D-particles. In the
first quantised (world-sheet) language, such fluctuations are also described
by Riemann surfaces of higher topology with appropriate Dirichlet boundary
conditions (c.f. fig.~\ref{fig:dbranes}). The plethora of Feynman diagrams in
higher order quantum field theory is replaced by a small set of world sheet
diagrams classified by moduli which need to be summed or integrated over
\cite{zwiebach}. In order to understand possible consequences for CPT due to
space-time foam we will have to characterise the latter. As a result, D-particles cross the brane world, and thereby
appear as foamy flashing on and off structures for an observer on the
brane. 
When low energy matter given by a closed string propagating in a $\left(  d+1\right)
$-dimensional space-time collides with a very massive D-particle (0-brane)
embedded in this space-time, the D-particle recoils as a result \cite{kogan}.
We shall consider the simple case of bosonic stringy matter coupling to
D-particles and so we can only discuss matters of principle and ignore issues
of stability. However we should note that an open string model needs to
incorporate for completeness, higher dimensional D-branes such as the D3
brane. This is due to the vectorial charge carried by the string owing to the
Kalb-Ramond field. Higher dimensional D-branes (unlike D-particles) can carry
the charge from the endpoints of open strings that are attached to them. For a
closed bosonic string model the inclusion of such D-branes is not imperative
(see figure \ref{fig:recoil}). The higher dimensional branes are not pertinent
to our analysis however. The current state of phenomenolgical modelling of the
interactions of D-particle foam with stringy matter will be briefly summarised
now. Since there are no rigid bodies in general relativity the recoil
fluctuations of the brane and their effective stochastic back-reaction on
space-time cannot be neglected. D-particle recoil in the
''tree approximation'' i.e. in lowest order in the string coupling $g_{s}$,
corresponds to the punctured disc or Riemann sphere approximation in open or
closed string theory, induces a non-trivial space-time metric. For closed
strings colliding with a heavy (non-relativistic) D-particle the metric has
the form \cite{mav2}
\begin{equation}
g_{ij}=\delta_{ij},\,g_{00}=-1,g_{0i}=\varepsilon\left(  \varepsilon
y_{i}+u_{i}t\right)  \Theta_{\varepsilon}\left(  t\right)  ,\;i=1,\ldots,d.
\label{recoilmetric}%
\end{equation}
where the suffix $0$ denotes temporal (Liouville) components and
\begin{eqnarray}
   \Theta_{\varepsilon}\left(  t\right) &=& \frac{1}{2\pi i}\int_{-\infty
}^{\infty}\frac{dq}{q-i\varepsilon}e^{iqt},\label{heaviside}\\
  u_{i} &=& \left(  k_{1}-k_{2}\right)  _{i}\;,\nonumber
\end{eqnarray}
with $k_{1}\left(  k_{2}\right)  $ the momentum of the propagating
closed-string state before (after) the recoil; $y_{i}$ are the spatial
collective coordinates of the D particle and $\varepsilon^{-2}$ is identified
with the target Minkowski time $t$ for $t\gg0$ after the collision~\cite{kmw}.
These relations have been calculated for non-relativistic branes where $u_{i}$
is small. Now for large $t,$ to leading order,%
\begin{equation}
g_{0i}\simeq\overline{u}_{i}\equiv\frac{u_{i}}{\varepsilon}\propto\frac{\Delta
p_{i}}{M_{P}} \label{recoil}%
\end{equation}
where $\Delta p_{i}$ is the momentum transfer during a collision and $M_{P}$
is the Planck mass (actually, to be more precise $M_{P}=M_{s}/g_{s}$, where
$g_{s}<1$ is the (weak) string coupling, and $M_{s}$ is a string mass scale);
so $g_{0i}$ is constant in space-time but depends on the energy content of the
low energy particle \cite{Dparticle}. Such a feature does not arise in
conventional approaches to space-time foam and will be important in our
formulation of one of the microscopic models that we will consider. In a D-particle recoil the velocity $u_{i}$ is random and in the absence of any particular prior knowledge it is natural to take a distribution which is gaussian with $0$ mean and variance $\sigma$. However there is a further uncertainty in $u_{i}$. viz. quantum-fluctuation aspects of $u^{i}$ about its classical trajectory. Going
to higher orders in perturbation theory of the quantum field theory at fixed
genus does not qualitatively alter the situation in the sense that the
equation for $u^{i}$ remains deterministic. The effect of string perturbation theory where higher genus surfaces are considered (on taking into account the sub-leading leading infrared divergences)is to produce diffusive behaviour of $u^{i}$. The resulting probability distribution $p\left(u \right)$ for $u$  (ignoring the vectorial index of $u$) is \cite{nonextensive}
\begin{equation}
p\left(u \right)\\
=\sqrt{\frac{15\alpha^{\prime}}{2\pi\left(  g_{s}^{2}\eta\left(  t\right)
+15\alpha^{\prime}\sigma^{2}\right)  }}\exp\left[  -\frac{15\alpha^{\prime}%
}{2\left(  g_{s}^{2}\eta\left(  t\right)  +15\alpha^{\prime}\sigma^{2}\right)
}u^{2}\right]  ~.\label{gaussian2}
\end{equation}
where%
\begin{equation}
\eta\left(  t\right)  =2t_{0}^{2}+3\left(  t+t_{0}\right)  ^{2}\sqrt{1+\frac
{t}{t_{0}}}-5t_{0}^{\frac{1}{2}}(t+t_{0})^{\frac{3}{2}}. \label{dispersion}%
\end{equation}
The interaction time
includes \emph{both} the time for capture and re-emission of the string by the
D-particle, as well as the time interval until the next capture, during string propagation.
In a generic situation, this time could be much larger than
the capture time, especially in dilute gases of D-particles, which include less than one D-particle per string ($\alpha^{3/2}$) volume. Indeed, as discussed
in detail in \cite{emnuncertnew}, using generic properties of strings
consistent with the space-time uncertainties~\cite{yoneya}, the capture and
re-emission time $t_{0}$, involves the growth of a stretched
string between the string state and the D-brane world (c.f.
fig.~\ref{fig:recoil}) and is found proportional to the incident string energy
$p^{0}$:
\begin{equation}
t_{0} \sim\alpha^{\prime}p^0  \ll \sqrt{\alpha ^{\prime}}~.
\end{equation}
From the knowledge of $p\left(u \right)$ we can extract ${var}(u)$ the variance of $u$
which is an important  input for our estimate of the
omega effect:
\begin{equation}\label{uvar}
{var}(u)=\frac{\eta\left(  t\right)  g_{0}^{2}%
+15\alpha^{\prime}\sigma^{2}}{15\alpha^{\prime}}.
\end{equation}
Further details and discussion on string perturbation theory leading to (\ref{uvar}) can be found in \cite{mavspinstat}.

\section{Recoil model for $\omega$-effect}
\label{sec:1}
For an observer on the brane world the crossing D-particles
will appear as twinkling space-time defects, i.e. microscopic space-time
fluctuations. This will give the four-dimensional brane world a ``D-foamy''
structure.The target space-time metric state, which is
close to being flat, can be represented schematically as a density matrix
\begin{equation}
\rho_{\mathrm{grav}}=\int d\,^{5}r\,\,f\left(  r_{\mu}\right)  \left|
g\left(  r_{\mu}\right)  \right\rangle \left\langle g\left(  r_{\mu}\,\right)
\right|  .\, \label{gravdensity}%
\end{equation}
\bigskip The parameters $r_{\mu}\,\left(  \mu=0,\ldots,5\right)  $ \ are
stochastic with a gaussian distribution $\,f\left(  r_{\mu}\,\right)  $
characterised by the averages%
\begin{equation}
\left\langle r_{\mu}\right\rangle =0,\;\left\langle r_{\mu}r_{\nu
}\right\rangle =\Delta_{\mu}\delta_{\mu\nu}\,\label{stoch}.
\end{equation}
The fluctuations experienced by the two entangled neutral mesons will be
assumed to be independent and $\Delta_{\mu}\sim O\left(  \frac{E^{2}}%
{M_{P}^{2}}\right)  $i.e. very small. As matter moves through the space-time
foam, assuming ergodicity, the effect of time averaging is assumed to
be equivalent to an ensemble average. As far as our present discussion is
concerned we will consider a semi-classical picture for the metric and so
$\left|  g\left(  r_{\mu}\right)  \right\rangle $ in \ref{gravdensity} will be
a coherent state.In order to address flavour oscillation
phenomena~\cite{NickSarben}~\cite{GabNick} the fluctuations
of each component of the metric tensor $g^{\alpha\beta}$ will not be
simply given by the simple recoil distortion (\ref{recoil}), but
instead will be taken to have a $2\times2$ (``flavour'') structure:
\begin{eqnarray}
  g^{00} &=& \left(  -1+r_{4}\right)  \mathsf{1}\nonumber \\
  g^{01} &=& g^{10}=r_{0}\mathsf{1}+ r_{1}\sigma_{1}+ r_{2}\sigma_{2}%
+r_{3}\sigma_{3}\label{metric} \\
  g^{11} &=& \left(  1+r_{5}\right)  \mathsf{1}\nonumber
\end{eqnarray}
The above parametrisation has been taken for simplicity and we will also
consider motion to be in the $x$- direction which is natural since the meson
pair moves collinearly in the Center-of-Mass frame.A metric with this type of structure is
compatible with the view that the D-particle defect is a ``point-like''
approximation for a compactified higher-dimensional brany black hole, whose no
hair theorems permit non-conservation of flavour. In the case of neutral
mesons the concept of ``flavour''
refers to either particle/antiparticle species or the two mass eigenstates,
by changing appropriately the relevant coefficients.
Space-time deformations of the form (\ref{metric}) and the associated
Hamiltonians (\ref{GenKG}) have been derived in the context of conformal field
theory~\cite{Dfoam,szabo2,nonextensive,mavspinstat} and details on the relevant derivations will
not be given here. We only mention that the variable $r_{1}$ in particular,
expresses a momentum transfer during the interaction of the (string) matter
state with the D-particle defect. In this sense, the off-diagonal metric
component $g_{01}$ can be represented as
\begin{equation}
g_{01} \sim u_{1} \label{umetric}%
\end{equation}
where $u_{1} = g_{s}\frac{\Delta k}{M_{s}}$ expresses the momentum transfer
along the direction of motion of the matter string (taken here to be the x
direction). In the above equation, $M_{s}/g_{s}$ is the mass of the
D-particle, which for weakly coupled strings with coupling $g_{s}$ is larger
than the string mass scale $M_{s}$. In order to address oscillation phenomena,
induced by D-particles, the fluctuations of each component of the metric
tensor are taken in \cite{bernabeu} to have a $2\times2$ (``flavour'')
structure, as in (\ref{metric}).

For the neutral Kaon system, the case of interest, $K_0 -{\overline K}_0$, is produced by a $\phi$-meson at rest,i.e. $K_0 -{\overline K}_0$ in their C.M. frame.
The CP eigenstates (on choosing a suitable
phase convention for the states $\left|  K_{0}\right\rangle $ \ and $\left|
\overline{K_{0}}\right\rangle $ ) are, in standard notation, $\left|  K_{\pm
}\right\rangle $ with
\begin{equation}
\left|  K_{\pm}\right\rangle =\frac{1}{\sqrt{2}}\left(  \left|  K_{0}%
\right\rangle \pm\left|  \overline{K_{0}}\right\rangle \right)  . \label{cp}%
\end{equation}
The mass eigensates $\left|  K_{S}\right\rangle $ and $\left|  K_{L}%
\right\rangle $ are written in terms of $\left|  K_{\pm}\right\rangle $ as%

\begin{equation}
\left|  K_{L}\right\rangle =\frac{1}{\sqrt{1+\left|  \varepsilon_{2}\right|
^{2}}}\left[  \left|  K_{-}\right\rangle \,+\varepsilon_{2}\left|
K_{+}\right\rangle \right]  \label{Klong}%
\end{equation}

and
\begin{equation}
\left|  K_{S}\right\rangle =\frac{1}{\sqrt{1+\left|  \varepsilon_{1}\right|
^{2}}}\left[  \left|  K_{+}\right\rangle \,+\varepsilon_{1}\left|
K_{+}\right\rangle \right]  . \label{Kshort}%
\end{equation}

In terms of the mass eigenstates
\begin{equation}
\left|  i\right\rangle =\mathcal{C}\left\{
\begin{array}
[c]{c}%
\left(  \left|  K_{L}\left(  \overrightarrow{k}\right)  \right\rangle \left|
K_{S}\left(  -\overrightarrow{k}\right)  \right\rangle -\left|  K_{S}\left(
\overrightarrow{k}\right)  \right\rangle \left|  K_{L}\left(  -\overrightarrow
{k}\right)  \right\rangle \right)  +\\
\omega\left(  \left|  K_{S}\left(  \overrightarrow{k}\right)  \right\rangle
\left|  K_{S}\left(  -\overrightarrow{k}\right)  \right\rangle -\left|
K_{L}\left(  \overrightarrow{k}\right)  \right\rangle \left|  K_{L}\left(
-\overrightarrow{k}\right)  \right\rangle \right)
\end{array}
\right\}  \label{CPTV}%
\end{equation}
where $\mathcal{C=}\frac{\sqrt{\left(  1+\left|  \varepsilon_{1}\right|
^{2}\right)  \left(  1+\left|  \varepsilon_{2}\right|  ^{2}\right)  }}%
{\sqrt{2}\left(  1-\varepsilon_{1}\varepsilon_{2}\right)  }$ \cite{Bernabeu}.
In the notation of two level systems (on suppressing the $\overrightarrow{k}$
label) we write%
\begin{eqnarray}
\left|  K_{L}\right\rangle  &=& \left|  \uparrow\right\rangle \\
\left|  K_{S}\right\rangle  &=& \left|  \downarrow\right\rangle .\nonumber
\end{eqnarray}
These will be our ``flavours''
and represent the two physical eigenstates, with masses $m_{1} \equiv m_{L}%
$, $m_{2} \equiv m_{S}$, with
\begin{equation}
\Delta m = m_{L} - m_{S} \sim3.48 \times10^{-15}~\mathrm{GeV}~. \label{deltam}%
\end{equation}
The stochastic variables $r_{\mu}$ ((\ref{stoch})) in
(\ref{metric}), are linked with the fluctuations of the D-particle recoil
velocity, by representing the latter as:
\begin{equation}
u_{1} \sim r g_{s} \frac{k}{M_{s}}%
\end{equation}
upon the above-mentioned technicality of considering flavour changes in
addition to the momentum transfer. Hence, the detailed discussion in the previous session on the
stochastic fluctuations of the recoil velocity about a zero average value,
translates into rewriting (\ref{stoch}) with variances (c.f. (\ref{uvar}))
\begin{equation}
\Delta_{\mu}\sim g_{0}^{2}\frac{t_{0}{}^{2}}{\alpha^{\prime}}
+\mathcal{O}(\sigma^{2})\sim g_{0}^{2}\left(
\frac{p^{0}}{M_{s}}\right)  ^{2}
+\mathcal{O}(\sigma^{2})~,\quad\mu=1,2\label{varmod}.%
\end{equation}
Here we considered the capture time $t_{0}\sim\alpha^{\prime}p^{0}$, with
$p^{0}$ the energy of the probe, as spanning the essential interaction time
for the initial entangled meson state \cite{nonextensive}.In this way we extrapolate the result
(\ref{gaussian2}) to times smaller than $\sqrt{\alpha^{\prime}}$. This is
acceptable, as long as such times are \emph{finite}. From a conformal field
theory point of view, this means that we consider the world-sheet scaling
parameter $1/\varepsilon^{2}\sim\mathrm{ln}(L/a)^{2}\sim t_{0}\ll\sqrt
{\alpha^{\prime}}$ for probe energies $p^{0}\ll M_{s}$.

The Klein-Gordon equation for a spinless neutral meson field $\Phi=\left(
\begin{array}
[c]{c}%
\phi_{1}\\
\phi_{2}%
\end{array}
\right)  $ with mass matrix $m=\frac{1}{2}\left(  m_{1}+m_{2}\right)
\mathsf{1}+$ $\frac{1}{2}\left(  m_{1}-m_{2}\right)  \sigma_{3}$ in a
gravitational background is
\begin{equation}
(g^{\alpha\beta}D_{\alpha}D_{\beta}-m^{2})\Phi=0 \label{KleinGordon}%
\end{equation}
where $D_{\alpha}$ is the covariant derivative. Since the Christoffel symbols
vanish for $a_{i}$ independent of space time the $D_{\alpha}$ coincide with
$\partial_{\alpha}$. Hence
\begin{equation}
\left(  g^{00}\partial_{0}^{2}+2g^{01}\partial_{0}\partial_{1}+g^{11}%
\partial_{1}^{2}\right)  \Phi-m^{2}\Phi=0. \label{KG2}%
\end{equation}
It is useful at this stage to rewrite the state $\left|  i\right\rangle $ in
terms of the mass eigenstates.

\bigskip The unnormalised state \ $\left|  i\right\rangle $ will then be an
\ example of an initial state
\begin{equation}
\left|  \psi\right\rangle = \left|  k,\uparrow\right\rangle ^{\left(  1\right)
}\left|  -k,\downarrow\right\rangle ^{\left(  2\right)  }-\left|
k,\downarrow\right\rangle ^{\left(  1\right)  }\left|  -k,\uparrow
\right\rangle ^{\left(  2\right)  }+ |\Delta\rangle
\end{equation}
with
\[
|\Delta\rangle = \xi\left|  k,\uparrow\right\rangle
^{\left(  1\right)  }\left|  -k,\uparrow\right\rangle ^{\left(  2\right)
}+\xi^{\prime}\left|  k,\downarrow\right\rangle ^{\left(  1\right)  }\left|
-k,\downarrow\right\rangle ^{\left(  2\right)  } \label{initstate}%
\]
where $\left|  K_{L}\left(  \overrightarrow{k}\right)  \right\rangle =\left|
k,\uparrow\right\rangle $ and we have taken $\overrightarrow{k}$ to have only
a non-zero component $k$ in the $x$-direction; superscripts label the two
separated detectors of the collinear meson pair, $\xi$ and $\xi^{\prime}$ are
complex constants and we have left the state $\left|  \psi\right\rangle $
unnormalised. The evolution of this state is governed by a hamiltonian
$\widehat{H}$
\begin{equation}
\widehat{H}=g^{01}\left(  g^{00}\right)  ^{-1}\widehat{k}-\left(
g^{00}\right)  ^{-1}\sqrt{\left(  g^{01}\right)  ^{2}{k}^{2}-g^{00}\left(
g^{11}k^{2}+m^{2}\right)  } \label{GenKG}%
\end{equation}
which is the natural generalisation of the standard Klein-Gordon hamiltonian
in a one particle situation. Moreover $\widehat{k}\left|  \pm k,\uparrow
\right\rangle =\pm k\left|  k,\uparrow\right\rangle $ together with the
corresponding relation for $\downarrow$.
We next note that the Hamiltonian interaction terms
\begin{equation}
\widehat{H_{I}}=-\left(  {r_{1}\sigma_{1}+r_{2}\sigma_{2}}\right)  \widehat
{k}\label{inter}%
\end{equation}
are the leading order contribution in the small parameters $r_{\mu}$ in the
Hamiltonian $H$ (\ref{GenKG}), since the corresponding variances $\sqrt
{\Delta_{\mu}}$ are small. The term (\ref{inter}), has been used in
\cite{bernabeu} as a \emph{perturbation} in the framework of non-degenerate
perturbation theory, in order to derive the \textquotedblleft
gravitationally-dressed\textquotedblright\ initial entangled meson states,
immediately after the $\phi$ decay. The result is:
\begin{equation}
  \left\vert {k,\uparrow}\right\rangle _{QG}^{\left(  1\right)  }\left\vert
{-k,\downarrow}\right\rangle _{QG}^{\left(  2\right)  }-\left\vert
{k,\downarrow}\right\rangle _{QG}^{\left(  1\right)  }\left\vert {-k,\uparrow
}\right\rangle _{QG}^{\left(  2\right)  }=|\Sigma\rangle + |\widetilde{\Delta}\rangle
\end{equation}
where
\[
|\Sigma\rangle=\left\vert {k,\uparrow}\right\rangle
^{\left(  1\right)  }\left\vert {-k,\downarrow}\right\rangle ^{\left(
2\right)  }-\left\vert {k,\downarrow}\right\rangle ^{\left(  1\right)
}\left\vert {-k,\uparrow}\right\rangle ^{\left(  2\right)  },
\]
\[
 |\widetilde{\Delta}\rangle ={\begin{array}{c}
                                \left\vert {k,\downarrow}\right\rangle ^{\left(  1\right)  }\left\vert
{-k,\downarrow}\right\rangle ^{\left(  2\right)  }\left(  {\beta^{\left(
1\right)  }-\beta^{\left(  2\right)  }}\right)  +\left\vert {k,\uparrow
}\right\rangle ^{\left(  1\right)  }\left\vert {-k,\uparrow}\right\rangle
^{\left(  2\right)  }\left(  {\alpha^{\left(  2\right)  }-\alpha^{\left(
1\right)  }}\right) \\
                                +\beta^{\left(  1\right)  }\alpha^{\left(  2\right)  }\left\vert
{k,\downarrow}\right\rangle ^{\left(  1\right)  }\left\vert {-k,\uparrow
}\right\rangle ^{\left(  2\right)  }-\alpha^{\left(  1\right)  }\beta^{\left(
2\right)  }\left\vert {k,\uparrow}\right\rangle ^{\left(  1\right)
}\left\vert {-k,\downarrow}\right\rangle ^{\left(  2\right)  }\label{entangl}%
                              \end{array}}
\]
and
\begin{equation}
\alpha^{\left(  i\right)  }=\frac{^{\left(  i\right)  }\left\langle
\uparrow,k^{\left(  i\right)  }\right\vert \widehat{H_{I}}\left\vert
k^{\left(  i\right)  },\downarrow\right\rangle ^{\left(  i\right)  }}%
{E_{2}-E_{1}}~,\quad\beta^{\left(  i\right)  }=\frac{^{\left(  i\right)
}\left\langle \downarrow,k^{\left(  i\right)  }\right\vert \widehat{H_{I}%
}\left\vert k^{\left(  i\right)  },\uparrow\right\rangle ^{\left(  i\right)
}}{E_{1}-E_{2}}~,~\quad i=1,2\label{qgpert2}%
\end{equation}
where the index $(i)$ runs over meson species (\textquotedblleft
flavours\textquotedblright) ($1\rightarrow K_{L},~2\rightarrow K_{S}$). The
reader should notice that the terms proportional to $\left(  {\alpha^{\left(
2\right)  }-\alpha^{\left(  1\right)  }}\right)  $ and $\left(  {\beta
^{\left(  1\right)  }-\beta^{\left(  2\right)  }}\right)  $ in (\ref{entangl})
generate $\omega$-like effects. We concentrate here for brevity and
concreteness in the strangeness conserving case of the $\omega$-effect in the initial decay of the $\phi$ meson~\cite{bernabeu1}, which
corresponds to $r_{i}\propto\delta_{i2}$. We should mention, however, that in
general quantum gravity does not have to conserve this quantum number, and in
fact strangeness-violating $\omega$-like terms are generated in this
problem through time evolution~\cite{bernabeu}.

We next remark that on averaging the density matrix over the random variables
$r_{i}$, which are treated as independent variables between the two meson
particles of the initial state (\ref{entangl}), we observe that only terms of
order $|\omega|^{2}$ will survive, with the order of $|\omega
|^{2}$ being
\begin{eqnarray}
|\omega|^{2} &=& \widetilde{\Delta}_{(1),(2)}\left(  \mathcal{O}\left(
\frac{1}{(E_{1}-E_{2})}(\langle\downarrow,k|H_{I}|k,\uparrow\rangle
)^{2}\right)  \right),\nonumber\\
  &=& \widetilde{\Delta}_{(1),(2)}\left(  \mathcal{O}\left(
\frac{\Delta_{2}k^{2}}{(E_{1}-E_{2})^{2}}\right)  \right)  \sim\widetilde
{\Delta}_{(1),(2)}\left(  \frac{\Delta_{2}k^{2}}{(m_{1}-m_{2})^{2}}\right)
\label{omegaorder}%
\end{eqnarray}
for the physically interesting case of non-relativistic Kaons in $\phi$
factories, in which the momenta are of order of the rest energies. The
notation $\widetilde{\Delta}_{(1),(2)}\left(  \dots\right)  $ above indicates
that one considers the differences of the variances $\Delta_{2}$ between the
two mesons $1$ - $2$, in that order.

The variances in our model of D-foam, which are due to quantum fluctuations of
the recoil velocity variables about the zero average (dictated by the imposed
requirement on Lorentz invariance of the string vacuum) are given by
(\ref{varmod}), with $p^{0}\sim m_{i}$ the \emph{energy} of the corresponding
individual (non-relativistic) meson state $(i)$ in the initial entangled state
(\ref{entangl}). It is important to notice that, on taking the difference of
the variance $\Delta_{2}$ between the mesons (1) and (2), the terms
proportional to the dispersion $\sigma^{2}$ in the initial recoil velocity
$u_{0}$ Gaussian distribution cancel out, since $\sigma^{2}$ is assumed
universal. In this sense, one is left with the contributions from the first
term of the right-hand-side of (\ref{varmod}), and thus we obtain the
following estimate \cite{nonextensive} for the square of the amplitude of the (complex) $\omega$-parameter:
\begin{equation}
|\omega|^{2}\sim g_{0}^{2}\frac{\left(  m_{1}^{2}-m_{2}^{2}\right)  }%
{M_{s}^{2}}\frac{k^{2}}{(m_{1}-m_{2})^{2}}
=\frac{m_{1}+m_{2}}{m_{1}-m_{2}}~\frac{k^{2}}{(M_{s}^{2}/g_{0}^{2})}
,\label{final}%
\end{equation}
where $M_{P}\equiv {M_{s}/g_{0}}$ represents the (average) quantum gravity scale, which may be taken to be
the four-dimensional Planck scale. In general, $M_{s}/g_{0}$ is the (average) D-particle
mass, as already mentioned. In the modern version of string theory, $M_{s}$ is
arbitrary and can be as low as a few TeV, but in order to have
phenomenologically correct string models with large extra dimensions one also
has to have in such cases very weak string couplings $g_{0}$, such that even
in such cases of low $M_{s}$, the D-particle mass $M_{s}/g_{0}$ is always
close to the Planck scale $10^{19}$ GeV. But of course one has to keep an open
mind about ways out of this pattern, especially in view of the string landscape.

The result (\ref{final}), implies, for neutral Kaons in a $\phi$ factory, for
which (\ref{deltam}) is valid, a value of: $|\omega|\sim 10^{-11}$,
which in the sensitive $\eta^{+-}$ bi-pion decay channel, is enhanced by three
orders of magnitude, as a result of the fact that the $|\omega|$
effect always appears in the corresponding observables~\cite{bernabeu1} in the
form $|\omega|/|\eta^{+-}|$, and the CP-violating parameter
$|\eta^{+-}|\sim10^{-3}$. Unfortunately, this value is still some two
orders of magnitude away from current bounds of the $\omega$-effect at,
or the projected sensitivity of upgrades of, the DA$\Phi$NE
detector~\cite{dafne}.

The estimate does not change much if one considers relativistic meson states.
Although in such a case the non-relativistic quantum mechanics formalism
leading to (\ref{final}) in the Kaon systems should strictly be replaced by an
appropriate relativistic treatment, nevertheless, one may still use the
expression (\ref{final}) in that case, in order to have a rough idea of the
order of magnitude of the effect in such relativistic cases. The major
difference in this case is the form of the Hamiltonian, which stems from the
expansion of the Dirac Hamiltonian for momenta $k\gg m_{i}$, $m_{i}$ the
masses. The quantities $E_{i}\sim k+\frac{m_{i}^{2}}{k^{2}},~i=1,2$ (due to
momentum conservation, assumed on average), and the capture times $t_{c}%
\sim\alpha^{\prime}E_{i}$. In such a case then, it is straightforward to
estimate $|{\omega}|^{2}$ as:
\begin{equation}
|\omega|^{2}\sim\widetilde{\Delta}_{(1),(2)}\left(  \Delta_{\mu}\frac{k^{2}%
}{(m_{1}-m_{2})^{2}}\right)  \sim\frac{m_{1}^{2}+m_{2}^{2}}{m_{1}^{2}%
-m_{2}^{2}}~\frac{g_{0}^{2}~k^{2}}{M_{s}^{2}}%
  ~,
\end{equation}
which does not lead to significant changes in the order of magnitude for
probes of energies up to 10 GeV. This completes our discussion on the
estimates of the $\omega$-effect in the initial entangled state of
two mesons in a meson factory. As discussed in \cite{bernabeu},
$\omega$-like terms can also be generated due to the time evolution
but we will not discuss this here.

\section{Alternative models of space-time foam}
\label{sec:1}
At this point the omega effect has been successfully generated through the string type decoherence model. Clearly the ability of the environment, in this case the D-particles, to change the flavour is important. For example models of fuzzy space-time foam embodied in non-commutative geometry capture effectively the non-commutative space-time commutation relations due to the presence (generally) of many D-branes \cite{zwiebach} in string theory and introduce non-locality. The resulting formalism,purely from the non-commutative structure, does not lead to change in flavours; for our model it was important that D-particle capture, recoil and emission of strings does not not conserve flavour. A simple calculation, within the framework of non-commutative geometry \cite{balachandran}, involving twisted tensor products implies a change of the EPR-type correlation of the CPT invariant form during time evolution (cf(\ref{CPTV})  with $\omega=0$) in the form of a small admixture of a term of the form $\left\vert {k,\uparrow}\right\rangle
^{\left(  1\right)  }\left\vert {-k,\downarrow}\right\rangle ^{\left(
2\right)  }+\left\vert {k,\downarrow}\right\rangle ^{\left(  1\right)
}\left\vert {-k,\uparrow}\right\rangle ^{\left(  2\right)  }$. This term is multiplied by the antisymmetric tensor $\theta^{\mu\nu}$ where $[x^{\mu},x^{\nu}]=i\theta^{\mu\nu}$. However this is not the correlation for the  $\omega$-effect. Hence within a string inspired framework it seems to be necessary to consider a space-time foam with a flavour changing mechanism. We will now consider a class of models of decoherence due to space-time foam which satisfies this criterion and is not directly related to string theory. It is based on a heuristic non-local effective theory approach to gravitational fluctuations. The non-locality arises because the fluctuation scale is taken to be intermediate between the Planck scale
and the low energy scale. Furthermore the dominant non-locality is assumed to
be bilocal, and, from similarities to quantum Brownian motion \cite{gardiner}%
,\cite{giulini}, plausible arguments can be given for a thermal bath model for
space-time foam \cite{garay}. The robustness or otherwise of the $\omega$-effect can be gauged from the predictions of this class of models.

Garay \cite{garay} has argued that the effect of non-trivial topologies
related to space-time foam and a non-zero minimum length can be modelled by a
field theory with non-local interactions on a flat background in terms of a
complete set of local functions $\left\{  h_{j}\left(  \phi,x\right)
\right\}  $ of the fields $\phi$ at a space-time point $x$. His argument is
based on general arguments related to problems of measurement \cite{giulini}.
Also, by considering the infinite redhshifting near the horizon for an
observer far away from the horizon of a black hole, Padmanabhan
\cite{padmanabhan} has argued that a foam consisting of virtual black holes
would magnify Planck scale physics for observers asymptotically far from the
horizon and thus an effective non-local field theory description would be
appropriate. The non-local action $I$ can be written in terms of a sum of
non-local terms $I_{n}$ where
\begin{equation}
I_{n}=\frac{1}{n!}\int dx_{1}\ldots\int dx_{n}\,\,f^{i_{1}\ldots i_{n}}\left(
x_{1},\ldots,x_{n}\right)  h_{i_{1}}\left(  x_{1}\right)  h_{i_{2}}\left(
x_{2}\right)  \ldots h_{i_{n}}\left(  x_{n}\right)  \label{nonlocal}%
\end{equation}
(where a summation convention for the indices has been assumed; the
$f^{i_{1}\ldots i_{n}}$ depend on relative co-ordinates such as $x_{1}-x_{2}$
and are expected to fall off for large separations, the scale of fluctuations
being $l_{\ast}>\frac{1}{M_{P}}$ ). Assuming a form of weak coupling
approximation it was further argued \cite{garay} that the retention of only
the $n=2$ term would be a reasonable approximation. Formally (i.e. ignoring
the validity of Euclidean to Minkowski Wick rotations) the resulting
non-locality can be written in terms of an auxiliary field $\varphi$ through
the functional identity
\[
\exp\left(  i\int dx_{1}\int dx_{2}\,\,f^{i_{1}i_{2}}\left(  x_{1}%
-x_{2}\right)  h_{i_{1}}\left(  x_{1}\right)  h_{i_{2}}\left(  x_{2}\right)
\right)
=\wp[\phi]
\]
where
\[
\wp[\phi]=\int d\varphi\exp\left( \begin{array}{c}
                -\int\int dx_{1}dx_{2}\,k_{i_{1}i_{2}}\left(
x_{1}-x_{2}\right)  \varphi^{i_{1}}\left(  x_{1}\right)  \varphi^{i_{1}%
}\left(  x_{2}\right) \\
               +i\int dx\,\varphi^{j}\left(  x\right)  h_{j}\left(
x\right)
             \end{array}
   \right)
\]
and $k_{i_{1}i_{2}}$ is the inverse of $f^{i_{1}i_{2}}$. The bilocality is
now represented as a local field theory for $\phi$ subjected to a stochastic
field $\varphi$. As shown in \cite{barenboim2} and \cite{loreti} this
stochastic behaviour results \cite{garay} in a master equation \ of the type
found in quantum Brownian motion, i.e. unitary evolution supplemented by
diffusion. There is no dissipation which might be expected from the
fluctuation diffusion theorem because the noise is classical. \ Since the
energy scales for typical experiments are much smaller than those associated
with gravitational quantum fluctuations Garay considered $f^{i_{1}i_{2}%
}\left(  x_{1}-x_{2}\right)  $ to be proportional to a Dirac delta function
and argued that the master equation was that for a thermal bath.

A standard approach to quantum Brownian motion of a particle is to have the
particle interact with a bath of quantum harmonic oscillators \cite{gardiner}. A
thermal field represents a bath about which there is minimal information
since only the mean energy of the bath is known, a situation which may be valid
also for space-time foam. In applications of quantum information it has been
shown that a system of two qubits (or two-level systems) initially in a
separable state (and interacting with a thermal bath) can actually be
entangled by such a single mode bath \cite{entanglement}. As the system
evolves the degree of entanglement is sensitive to the initial state. The
close analogy between two-level systems and neutral meson systems, together
with the modelling by a phenomenological thermal bath of space-time foam,
makes the study of thermal master equations an intriguing one for the
generation of $\omega$-terms.The hamiltonian $\mathcal{H}$ representing the
interaction of \ two such two-level `atoms' with a single mode thermal field
\cite{Jaynes} is
\begin{equation}
\mathcal{H}=\nu a^{\dag}a+\frac{1}{2}\Omega\sigma_{3}^{\left(  1\right)
}+\frac{1}{2}\Omega\sigma_{3}^{\left(  2\right)  }+\gamma\sum_{i=1}^{2}\left(
a\sigma_{+}^{\left(  i\right)  }+a^{\dag}\sigma_{-}^{\left(  i\right)
}\right)  \label{thermal}%
\end{equation}
where $a$ is the annihilation operator for the mode of the thermal field and
the $\sigma$'s are again the Pauli matrices for the 2-level systems (using the
standard conventions).The superscripts label the particle.The harmonic oscillator operators $a$ and $a^{\dag}$
satisfy
\begin{equation}
\left[  a,a^{\dag}\right]  =1,\left[  a^{\dag},a^{\dag}\right]  =\left[
a,a\right]  =0. \label{commutation}%
\end{equation}
The $\mathcal{H}$ here (commonly known as the
Jaynes-Cummings hamiltonian \cite{Jaynes}) is quite different from the $\mathcal{H}$ of the
D-particle foam of the last section. It explicitly incorporates the back
reaction or entanglement between system and bath. There are no classical stochastic terms
at this level of description and also $\mathcal{H}$ is \emph{not} separable. In the
D-particle foam model the lack of separabillity came solely from the entangled
nature of the initial unperturbed state. This is also in contrast with the
Lindlblad model. The thermal master equation comes
from tracing over the oscillator degrees of freedom. We will however consider
the dynamics before tracing over the bath because, although the thermal bath
idea has a certain intuitive appeal, it cannot claim to be rigorous and so for
attempts to find models for the $\omega$-effect it behoves us to entertain
also deviations from the thermal bath state of the reservoir.
An important feature of (\ref{thermal}) is the block structure of subspaces that are left invariant by
$\mathcal{H}$. It is straightforward to show that the family of invariant
irreducible spaces $\mathcal{E}_{n}$ may be defined by $\left\{  \left\vert
e_{i}^{\left(  n\right)  }\right\rangle ,i=1,\ldots,4\right\}  $ where ( in
obvious notation, with $n$ denoting the number of oscillator quanta)
\begin{equation}
\left\vert e_{1}^{\left(  n\right)  }\right\rangle   \equiv\left\vert
\uparrow^{\left(  1\right)  },\,\uparrow^{\left(  2\right)  },n\right\rangle
,\label{equation1}\\
\left\vert e_{2}^{\left(  n\right)  }\right\rangle  \equiv\left\vert
\uparrow^{\left(  1\right)  },\,\downarrow^{\left(  2\right)  }%
,n+1\right\rangle ,\nonumber\\
\left\vert e_{3}^{\left(  n\right)  }\right\rangle   \equiv\left\vert
\downarrow^{\left(  1\right)  },\,\uparrow^{\left(  2\right)  }%
,n+1\right\rangle ,\nonumber
\end{equation}
and
\[
\left\vert e_{4}^{\left(  n\right)  }\right\rangle \equiv\left\vert
\downarrow^{\left(  1\right)  },\,\downarrow^{\left(  2\right)  }%
,n+2\right\rangle .
\]
The total space of states is a direct sum of the $\mathcal{E}_{n}$ for
different $n$. We will write $\mathcal{H}$ as $\mathcal{H}_{0}+\mathcal{H}%
_{1}$ where
\begin{equation}
\mathcal{H}_{0}=\nu a^{\dag}a+\frac{\Omega}{2}\left(  \sigma_{3}^{\left(
1\right)  }+\sigma_{3}^{\left(  2\right)  }\right)  \label{hamiltonian0}%
\end{equation}
and
\begin{equation}
\mathcal{H}_{1}=\gamma^{\prime}\sum_{i=1}^{2}\left(  a\sigma_{+}^{\left(
i\right)  }+a^{\dag}\sigma_{-}^{\left(  i\right)  }\right)  .
\label{hamiltonianI}%
\end{equation}
$n$ is a quantum number and gives the effect of the random environment. In our
era the strength $\gamma$ of the coupling with the bath is weak. We expect
heavy gravitational degrees of freedom and so $\Omega\gg\nu$. It is possible
to associate both thermal and highly non-classical density matrices with the
bath state.\ Nonethelss because of the block structure it can be shown
completely non-perturbatively that the $\omega$ contribution to the density
matrix is absent \cite{sarkar2}.

\bigskip We will calculate the stationary states in $\mathcal{E}_{n}$, using
degenerate perturbation theory, where appropriate. We will be primarily
interested in the dressing of the degenerate states $\left|  e_{2}^{\left(
n\right)  }\right\rangle $ and $\left|  e_{3}^{\left(  n\right)
}\right\rangle $ because it is these which contain the neutral meson entangled
state . In 2nd order perturbation theory the dressed states are
\begin{equation}
\left|  \psi_{2}^{\left(  n\right)  }\right\rangle =\left|  e_{2}^{\left(
n\right)  }\right\rangle -\left|  e_{3}^{\left(  n\right)  }\right\rangle
+O\left(  \gamma^{\prime\,3}\right)  \label{dressed1}%
\end{equation}
with energy $E_{2}^{\left(  n\right)  }=\left(  n+1\right)  \nu+O\left(
\gamma^{\prime\,3}\right)  $ and
\begin{equation}
\left|  \psi_{3}^{\left(  n\right)  }\right\rangle =\left|  e_{2}^{\left(
n\right)  }\right\rangle +\left|  e_{3}^{\left(  n\right)  }\right\rangle
+O\left(  \gamma^{\prime\,3}\right)  \label{dressed2}%
\end{equation}
with energy $E_{3}^{\left(  n\right)  }=\left(  n+1\right)  \nu+\frac
{2\gamma^{2}}{\Omega-\nu}+O\left(  \gamma^{\prime\,3}\right)  $. It is
$\left|  \psi_{2}^{\left(  n\right)  }\right\rangle $ which can in principle
give the $\omega$-effect. More precisely we would construct the state
$Tr\left(  \left|  \psi_{2}^{\left(  n\right)  }\right\rangle \left\langle
\psi_{2}^{\left(  n\right)  }\right|  \rho_{B}\right)  $ where $\rho_{B}$ is
the bath density matrix (and has the form $\rho_{B}=\sum_{n,m}p_{nn^{\prime}%
}\left|  n\right\rangle \left\langle n^{\prime}\right|  $ for suitable choices
of $p_{nn^{\prime}}$). To this order of approximation $\left|  \psi
_{2}^{\left(  n\right)  }\right\rangle $ cannot generate the $\omega$-effect
since there is no admixture of $\left|  e_{1}^{\left(  n\right)
}\right\rangle $ and $\left|  e_{4}^{\left(  n\right)  }\right\rangle $.
However it is a priori possible that this may change when higher orders in
$\gamma$. We can show that $\left|  \psi_{2}^{\left(  n\right)  }\right\rangle
=\left|  e_{2}^{\left(  n\right)  }\right\rangle -\left|  e_{3}^{\left(
n\right)  }\right\rangle $ to \textit{all orders} in $\gamma$ by directly
considering the hamiltonian matrix $\mathcal{H}^{\left(  n\right)  }$ for
$\mathcal{H}$ within $\mathcal{E}_{n}$; it is given by
\begin{equation}
\mathcal{H}^{\left(  n\right)  }=\left(
\begin{array}
[c]{cccc}%
\Omega+n\nu & \gamma\sqrt{n+1} & \gamma\sqrt{n+1} & 0\\
\gamma\sqrt{n+1} & \left(  n+1\right)  \nu & 0 & \gamma\sqrt{n+2}\\
\gamma\sqrt{n+1} & 0 & \left(  n+1\right)  \nu & \gamma\sqrt{n+2}\\
0 & \gamma\sqrt{n+2} & \gamma\sqrt{n+2} & \left(  n+2\right)  \nu-\Omega
\end{array}
\right)  . \label{matrix}%
\end{equation}
We immediately notice that
\begin{equation}
\mathcal{H}^{\left(  n\right)  }\left(
\begin{array}
[c]{c}%
0\\
1\\
-1\\
0
\end{array}
\right)  =\left(  n+1\right)  \nu\left(
\begin{array}
[c]{c}%
0\\
1\\
-1\\
0
\end{array}
\right)  \label{eigenvectoreqn}%
\end{equation}
and so, to all orders in $\gamma$, the environment does \textit{not} dress the
state of interest to give the $\omega$-effect; clearly this is independent of
any choice of $\rho_{B}$.

As we noted this block structure of the Hilbert space is a consequence of the
structure of $\mathcal{H}_{1}$ which is commonly called the hamiltonian in the
`rotating wave' approximation\cite{PikeSark}. (This nomenclature arises within
the context of quantum optics where this model is used extensively.) It is
natural to expect that, once the rotating wave approximation is abandoned, our
conclusion would be modified. We shall examine whether this expectation
materialises by adding to $\mathcal{H}$ a `non-rotating wave' piece
$\mathcal{H}_{2}$
\begin{equation}
\mathcal{H}_{2}=\gamma\sum_{i=1}^{2}\left(  a\sigma_{-}^{\left(  i\right)
}+a^{\dag}\sigma_{+}^{\left(  i\right)  }\right)  . \label{nonrotwave}%
\end{equation}
$\mathcal{H}_{2}$ does not map $\mathcal{E}_{n}$ into $\mathcal{E}_{n}$ as can
be seen from
\[
\mathcal{H}_{2}\left\vert e_{2}^{\left(  n\right)  }\right\rangle
=\gamma^{\prime}\left(  \sqrt{n+2}\,\left\vert e_{1}^{\left(  n+2\right)
}\right\rangle +\sqrt{n+1}\left\vert e_{4}^{\left(  n-2\right)  }\right\rangle
\right)  .
\]
We have noted that $\left\vert e_{3}^{\left(  n\right)  }\right\rangle
-\left\vert e_{2}^{\left(  n\right)  }\right\rangle $ is an eigenstate of
\ $\mathcal{H}_{0}+\mathcal{H}_{1}$. However the perturbation $\mathcal{H}%
_{2}$ \textit{annihilates} $\left\vert e_{3}^{\left(  n\right)  }\right\rangle
-\left\vert e_{2}^{\left(  n\right)  }\right\rangle $ and so this eigenstate
does not get dressed by $\mathcal{H}_{2}$. Hence the $\omega$ effect cannot be
rescued by moving away from the rotating wave scenario.

We should comment that a variant of the bath model based on non-bosonic
operators $a$ can be considered. \ This departs somewhat from the original
philosophy of Garay but may have some legitimacy from analysis in M-theory.
From an infinite momentum frame study it was suggested that D-particles may
satisfy infinite statistics\cite{InfStat}. Of course these D-particles would
be light contrary to the heavy D-particles that we have been considering and
so we will consider a more general deformation of the usual statistics.. If
flavour changes can arise due to D-particle interactions then it may be
pertinent to reinterpret the hamiltonian in this way. Generalised statistics
is characterised conventionally by a c-number deformation\cite{macfarlane}
parameter $q$ with $q=1$ giving bosons. Although we have as yet not made a
general analysis of such deformed algebras, we will consider the following
particular deformation of the usual harmonic oscillator algebra:
\begin{equation}
a_{q}a_{q}^{\dagger}-q^{-1}a_{q}^{\dagger}a_{q}=q^{N_{q}}\label{qdeform}%
\end{equation}
where the $q$-operators $a_{q},a_{q}^{\dagger}$ and $N_{q}$ act on a Hilbert
space with a\ denumerable orthonormal basis  $\left\{  \left\vert
n\right\rangle _{q},n=0,1,2,\ldots\right\}  $ as follows:%
\begin{equation}
a_{q}^{\dagger}\left\vert n\right\rangle _{q}   =\left[  n+1\right]
^{1/2}\left\vert n+1\right\rangle _{q},\nonumber\\
a_{q}\left\vert n\right\rangle _{q} =\left[  n\right]  ^{1/2}\left\vert
n-1\right\rangle _{q},\label{deformedHilbertspace}\\
N_{q}\left\vert n\right\rangle _{q} =n\left\vert n\right\rangle
_{q}\nonumber
\end{equation}
and
\begin{equation}
\left[  x\right]  \equiv\frac{q^{x}-q^{-x}}{q-q^{-1}}.
\end{equation}
Clearly when $q=1$ the boson case is recovered. We can consider the $q$
generalisation of $\mathcal{H}$.\ The resulting $q$ operator is
\begin{equation}
\mathcal{H}_{q}=\hbar\nu a_{q}^{\dag}a_{q}+\frac{1}{2}\hbar\Omega\sigma
_{3}^{\left(  1\right)  }+\frac{1}{2}\hbar\Omega\sigma_{3}^{\left(  2\right)
}+\hbar\gamma\sum_{i=1}^{2}\left(  a_{q}\sigma_{+}^{\left(  i\right)  }%
+a_{q}^{\dag}\sigma_{-}^{\left(  i\right)  }\right)  \label{qHamiltonian}%
\end{equation}
and there is still the natural $q$ analogue $\mathcal{E}_{n}^{\left(
q\right)  }$ of the subspace $\mathcal{E}_{n}$ which is invariant under
$\mathcal{H}_{q}$ and spanned by the basis
\begin{equation}
\left\vert e_{1}^{\left(  n\right)  }\right\rangle _{q}  \equiv\left\vert
\uparrow^{\left(  1\right)  },\,\uparrow^{\left(  2\right)  }\right\rangle
\left\vert n\right\rangle _{q},\\
\left\vert e_{2}^{\left(  n\right)  }\right\rangle _{q}   \equiv\left\vert
\uparrow^{\left(  1\right)  },\,\downarrow^{\left(  2\right)  }\right\rangle
\left\vert n+1\right\rangle _{q},\\
\left\vert e_{3}^{\left(  n\right)  }\right\rangle _{q} \equiv\left\vert
\downarrow^{\left(  1\right)  },\,\uparrow^{\left(  2\right)  }\right\rangle
\left\vert n+1\right\rangle _{q},
\end{equation}
and
\begin{equation}
\left\vert e_{4}^{\left(  n\right)  }\right\rangle _{q}\equiv\left\vert
\downarrow^{\left(  1\right)  },\,\downarrow^{\left(  2\right)  }\right\rangle
\left\vert n+2\right\rangle _{q}.
\end{equation}
A very similar calculation to the bosonic case yields
\begin{equation}
\mathcal{H}_{q}^{\left(  n\right)  }=\left(
\begin{array}
[c]{cccc}%
\Omega+\left[  n\right]  \nu & \gamma\sqrt{\left[  n+1\right]  } & \gamma
\sqrt{\left[  n+1\right]  } & 0\\
\gamma\sqrt{\left[  n+1\right]  } & \left[  n+1\right]  \nu & 0 & \gamma
\sqrt{\left[  n+2\right]  }\\
\gamma\sqrt{\left[  n+1\right]  } & 0 & \left[  n+1\right]  \nu & \gamma
\sqrt{\left[  n+2\right]  }\\
0 & \gamma\sqrt{\left[  n+2\right]  } & \gamma\sqrt{\left[  n+2\right]  } &
\left[  n+2\right]  \nu-\Omega
\end{array}
\right)  .\label{qdeformedinteraction}%
\end{equation}
Again the vector $\left\vert e_{2}^{\left(  n\right)  }\right\rangle
_{q}-\left\vert e_{3}^{\left(  n\right)  }\right\rangle _{q}$ is an
eigenvector with eigenvalue $\left[  n+1\right]  \nu$ . Hence this model using
generalised statistics based on the Jaynes-Cummings framework also does not
lead to the $\omega$-effect.

One cannot gainsay that other more complicated models of `thermal' baths may
display the $\omega$-effect but clearly a rather standard model rejects quite
emphatically the possibility of such an effect. This by itself is very
interesting. \ It shows that the $\omega$-effect is far from a generic
possibility for space-time foams. Just as it is remarkable that various
versions of the paradigmatic `thermal' bath as well as entanglement
associated with non-commutative geometry cannot accommodate the $\omega
$-effect, it is also remarkable that the D-particle foam model manages to do so
very simply.

%
%
\section{Conclusions}

{} In this work we have discussed in detail primarily two classes of space-time foam models, which
may characterise realistic situations of the (still elusive) theory of quantum
gravity. They both involve non-unitary evolution. Relevant to this is, as mentioned earlier,  standard models of non-commutative geometry, even though they incorporate fuzziness of space-time, are unable to reproduce the omega effect. In one of our models, inspired by non-critical string theory, string
mattter on a brane world interacts with D particles in the bulk. Recoil of the
heavy D particles owing to interactions with the stringy matter produces a
gravitational distortion which has a backreaction on the stringy matter. This
distortion, consistent with a logarithmic conformal field theory algebra,
depens on the recoil velocity of the D-particle. It is modelled by a
stochastic metric and consequently can affect the different mass eigenstates
of neutral mesons.There are two contributions to the stochasticity. One is due
to the stochastic aspects of the recoil velocity found in low order loop
amplitudes of open or closed strings ( performed explicitly in the bosonic
string theory model) while the other is the leading behaviour of an infinite
resummation of the subleading order infrared divergences in loop perturbation
theory. The former is assumed to fluctuate randomly, with a dispersion which
is viewed as a phenomenological parameter. The latter for long enough capture
times of the D-particle with stringy matter gives a calculable dispersion
which dominates the phenomenological dispersion for the usual expectations of
the quantum gravity scale. This gives a prediction for the order of magnitude
of the CPTV $\omega$-like effect in the initial entangled state of two neutral
mesons in a meson factory based on stationary (non degenerate) perturbation
theory for the gravitational dressing of the correlated meson state.The
Klein-Gordon hamiltonian in the induced stochastic gravitiational field was
used for the calculation of the dressing.The order of magnitude of $\omega$
may not be far from the sensitivity of immediate future experimental
facilities, such as a possible upgrade of the DA$\Phi$NE detector or a B-meson factory.

This causes a CPTV $\omega$-like effect in the initial entangled state of two
neutral mesons in a meson factory, of the type conjectured in \cite{Bernabeu}.
Using stationary perturbation theory it was possible to give an order of
magnitude estimate of the effect: the latter is momentum dependent, and of an
order which may not be far from the sensitivity of immediate future
experimental facilities, such as a possible upgrade of the DA$\Phi$NE detector
or a B-meson factory. A similar effect, but with a sinusoidal time dependence,
and hence experimentally disentanglable from the initial-state effect , is
also generated in this model of foam by the evolution of the system.

A second model of space-time foam, that of a thermal bath of gravitational
degrees of freedom, is also considered in our work, which, however, does not
lead to the generation of an $\omega$-effect. Compared to the initial
treatment of thermal baths further significant generalisations have been
considered here. In the initial treatment (using a model derived in the
rotating wave approximation) the structure of the hamiltonian matrix was block
diagonal, the blocks being invariant 4-dimensional subspaces.The real
significance of the block diagonal structure was unclear for the absence of the
omega effect in this model and has remained a matter for debate. We have
analyzed the model ( now without the rotating wave approximation) and to \ our
surprise the omega effect is still absent. Furthermore we have also considered
$q$ oscillators for the heat bath modes since from Matrix theory we know that
energetic D-particles satisfy $q$ statistics. For the model of $q$ statistics
that we have adopted there is again no $\omega$-effect in the $q$ version of
the thermal bath.

It is interesting to continue the search for other (and more) realistic models
of quantum gravity, either in the context of string theory or in alternative
approaches, such as loop quantum gravity, exhibiting intrinsic CPT Violating
effects in sensitive matter probes. Detailed analyses of global data in
relation to CPTV, including very sensitive probes such as high energy
neutrinos, is a good way forward.

{ \ }

\begin{acknowledgements}
I would like to thank the organisers of the SPIN-STAT $2008$ Conference (Trieste, Italy, October 21-25 2008) for the opportunity to present results from the current work. It is a pleasure to acknowledge partial support by the European Union through
the FP6 Marie Curie Research and Training Network UniverseNet (MRTN-CT-2006-035863).
\end{acknowledgements}



\end{document}